\documentclass[letterpaper, 10 pt, conference]{ieeeconf}  

\IEEEoverridecommandlockouts                              

\usepackage{amsmath} 
\usepackage{amssymb}  
\usepackage{commath}
\usepackage{soul,color}
\let\labelindent\relax
\usepackage[ruled,vlined]{algorithm2e}
\usepackage{enumitem}
\usepackage{graphicx}
\usepackage{mathptmx}
\usepackage{flushend}

\newtheorem{theorem}{Theorem}

\newtheorem{remark}{Remark}

\usepackage{color}

\usepackage{lscape}

\title{\LARGE \bf
State Estimation for a Zero-Dimensional Electrochemical Model \\ of Lithium-Sulfur Batteries
}

\author{Zhijia Huang$^{a,1}$, Dong Zhang$^{a,2}$, Luis D. Couto$^{3,b}$, Quan-Hong Yang$^{1,4}$, and Scott J. Moura$^{1,2}$ 
\thanks{$^{a}$Z. Huang and D. Zhang contributed equally to this work as first authors.}%
\thanks{\textcolor{black}
{$^{b}$Luis D.\ Couto would like to thank the Wiener-Anspach Foundation for its financial support.}}%
\thanks{$^{1}$Zhijia Huang, Quan-Hong Yang and Scott J. Moura is with Tsinghua-Berkeley Shenzhen Institute (TBSI), Tsinghua Shenzhen International Graduate School, Tsinghua University, Shenzhen, 518055,
China. {\tt\small hzj19@berkeley.edu}}%
\thanks{$^{2}$Dong Zhang and Scott J. Moura is with Department of Civil and Environmental Engineering, University of California, Berkeley, California, 94720,
USA. {\tt\small \{dongzhr,smoura\}@berkeley.edu}}%
\thanks{$^{3}$Luis D. Couto is with Department of Control Engineering and System Analysis, Universit\'e Libre de Bruxelles, B-1050 Brussels, Belgium. {\tt\small lcoutome@ulb.ac.be}}%
\thanks{$^{4}$Quan-Hong Yang is with State Key Laboratory of Chemical Engineering, School of Chemical Engineering and Technology, Tianjin University, Tianjin, 300072,
China. {\tt\small  qhyangcn@tju.edu.cn}}%
}

\begin{document}

\maketitle
\thispagestyle{empty}
\pagestyle{empty}

\begin{abstract}

Lithium-sulfur (Li-S) batteries have become one of the most attractive alternatives over conventional Li-ion batteries due to their high theoretical specific energy density (2500 Wh/kg for Li-S vs. $\sim$250 Wh/kg for Li-ion). Accurate state estimation in Li-S batteries is urgently needed for safe and efficient operation. To the best of the authors' knowledge, electrochemical model-based observers have not been reported for Li-S batteries, primarily due to the complex dynamics that make state observer design a challenging problem. In this work, we demonstrate a state estimation scheme based on a zero-dimensional electrochemical model for Li-S batteries. The nonlinear differential-algebraic equation (DAE) model is incorporated into an extend Kalman filter. This observer design estimates both differential and algebraic states that represent the dynamic behavior inside the cell, from voltage and current measurements only. The effectiveness of the proposed estimation algorithm is illustrated by numerical simulation results. Our study unlocks how an electrochemical model can be utilized for practical state estimation of Li-S batteries.

\end{abstract}

\section{Introduction}
Lithium ion batteries (LIBs) have been regarded as one of the most successful rechargeable battery systems in view of their excellent safety, high energy density, and long cycling performance. With these merits, LIBs have occupied a dominating market share in electric vehicles (EVs), large-scale energy storage systems and portable electronics. However, they still fail to meet the increasing demand for high energy density storage applications due to their limited capacity and low energy density. Consequently, there is a strong incentive to develop next generation battery systems with much higher energy densities \cite{bruce2012li,lin2017reviving}. Lithium-sulfur (Li-S) batteries, with abundant sulfur as the cathode material, are considered a promising candidate because of their high theoretical energy density, environmental friendliness and low cost. Unlike the ``insertion'' mechanism in LIBs, a multi-electron electrochemical redox reaction takes place in Li-S battery systems, where Li ions react with sulfur to generate lithium sulfide (Li$_2$S) at the end discharge. This achieves a much higher energy density of 2500 Wh$\cdot$kg$^{-1}$, which is almost 10 times higher than LIBs and would enable electrified aircraft and long-haul trucks. However, their practical use at large scales is hindered by several challenges. These include limited utilization of sulfur, complex reaction kinetics, and severe shuttling of soluble lithium polysulfides (LiPSs), all of which lead to poor cycling performance and fast capacity decay \cite{zhang2020optimized}. Tremendous efforts have been devoted to address these issues and promote the practical demonstration of Li-S batteries. Therefore, as a potential candidate for automotive applications, it is imperative to develop an advanced battery management system (BMS) \cite{chaturvedi2010algorithms} that implements real-time control and estimation algorithms for their practical use in the near future.

The techniques for state estimation in commercial LIBs are well-established, and the two straightforward categories are ``Coulomb counting'' \cite{xiong2017critical} and open-circuit voltage (OCV) measurement \cite{xing2014state}. However, these existing estimation algorithms cannot be directly applied to Li-S batteries due to their unique features, namely the complex reaction chemistry and the ``shuttle effect'' \cite{zhang2020optimized}. For instance, the Coulomb counting method is limited by the relatively high self-discharge behavior \cite{ryu2006self} and the high dependence of capacity on duty cycle and applied current profile \cite{ryu2009investigation}. Moreover, the open-circuit voltage in Li-S cells cannot be used as an indicator of state of charge (SOC) over the whole discharge range due to its unique shape. The typical OCV-SOC curve of Li-S batteries can be divided into two plateaus, where the low plateau has a flat region that reduces the observability of the system and hinders the application of OCV-based techniques \cite{fotouhi2017lithium,propp2017kalman}. 

Model-based estimation techniques have been widely employed in LIBs \cite{sepasi2014novel,fleischer2014line,bartlett2015electrochemical,moura2016battery,zhang2019battery}. In contrast, there are just a few similar studies for Li-S batteries due to the complex models used to characterize their electrochemical features. The Li-S battery models presented in literature can be classified into two main categories: the electrical equivalent circuit models (ECMs) and the physics-based electrochemical models. The proposed ECMs can reproduce the discharging behavior of Li-S batteries with current and temperature dependent parameters  \cite{knap2015electrical,propp2016multi}. Their intuitive structure and relatively low computational demands are suitable for battery state estimation and control. However, the electrochemical dynamics as well as internal electrochemical states cannot be accurately modelled. On the other hand, great efforts have been put on electrochemical models to understand the mechanisms inside the Li-S cells, including electrochemical reactions, shuttle effect and precipitation/dissolution processes \cite{kumaresan2008mathematical,neidhardt2012flexible,thangavel2016microstructurally,fronczek2013insight,hofmann2014mechanistic}. Due to the complex reaction pathways and various state variables of sulfur species involved, even for one-dimensional models \cite{fronczek2013insight,hofmann2014mechanistic}, they still require a large number of physical and chemical parameters as well as significant computational effort. Recently, reduced-order electrochemical models, especially zero-dimensional models \cite{marinescu2016zero}, have shown their advantages in accurately predicting the electrochemical dynamics with relatively low computational power, which provides a suitable tool to evaluate the performance of Li-S cells in various applications.  

To date, only few studies in the literature have demonstrated state estimation of Li-S batteries, all of which use ECMs. In \cite{propp2017kalman}, three recursive Bayesian state estimators, i.e., extended Kalman filter (EKF), unscented Kalman filter (UKF) and particle filter (PF), based on ECMs have been proposed to estimate SOC according to a combination of Coulomb counting and voltage response of the Li-S cell. However, they showed slow convergence if the initial conditions are unknown and became less accurate with variation of current profiles. To increase the robustness and accuracy, a ``behavioral form'' of the dual EKF with online parameter identification was then introduced for estimating the SOC and state of health (SOH) of Li-S batteries \cite{knap2018concurrent,propp2019improved}. 

The estimation of the various sulfur species is a crucial aspect in Li-S batteries since they determine the state-of-charge and state-of-health. 
Therefore, estimation offers a great potential to gain insight into the electrochemical mechanisms. Moreover, a zero-dimensional electrochemical model-based estimator enables practical applications. However, to the authors’ best knowledge, no previous work focuses on state estimation with electrochemical models for Li-S batteries, which is particularly challenging due to multiple technical reasons. First, the electrochemical models are generally governed by complex differential-algebraic equations. The measurable signals, namely cell current and voltage, are nonlinear in the system states and parameters. Second, the equilibrium potential of the low plateau of Li-S cells has a flat region that reduces the sensitivity of the output voltage with respect to the system states, rendering weak local observability. Finally, the analysis and estimator design tools for nonlinear DAEs have not been well understood for battery estimation problems. In light of aforementioned research gaps, in this paper we address these challenges by proposing a state estimation scheme for a reduced electrochemical model using voltage and current measurements only. The contributions of this work are summarized as follows:
\begin{enumerate}
    \item This is the first attempt in the literature to exploit reduced-order electrochemical models for state estimation in Li-S cells.
    \item An analysis of the local observability of the zero-dimensional Li-S battery model is provided using DAE techniques.
    \item We employ an extended Kalman filter for a DAE system to estimate the time evolution of sulfur species as well as reaction kinetics during the battery discharge process, enabling an unprecedented level of real-time monitoring for Li-S cells.  
\end{enumerate}

The reminder of this paper is organized as follows. The zero-dimensional electrochemical model is introduced in Section~\ref{s:mod}. In Section \ref{s:observability}, the local observability of the nonlinear DAE system is analyzed. An EKF algorithm for the state estimation of DAE systems is proposed in Section \ref{s:obs}. Simulation results and discussion are provided in Section \ref{s:sim}, followed by conclusions in Section \ref{s:conclusion}.

\begin{figure}[t]
	\centering
	\includegraphics[trim = 4.5mm 2mm 9mm 2mm, clip, width=\linewidth]{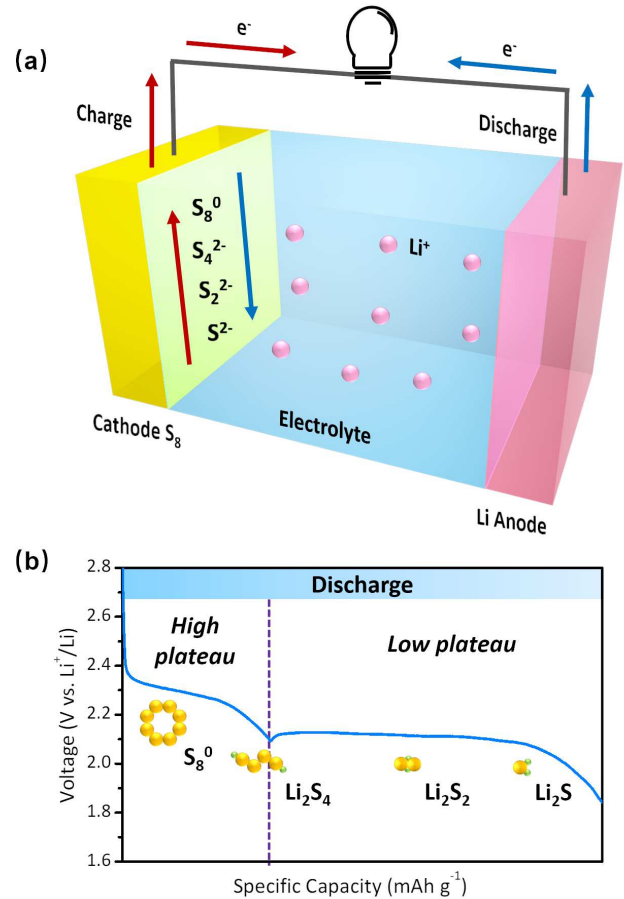}
	\caption{(a) A schematic illustration of the Li-S batteries. (b) Galvanostatic discharge profile and the typical sulfur species in each plateau.}
	\label{fig:LiS-Batt}
\end{figure}

\section{Li-S Battery Model} \label{s:mod}

Fig. \ref{fig:LiS-Batt}(a) shows the schematic of a Li-S cell composed of a sulfur cathode and a lithium metal anode. In a typical discharge process, elemental sulfur ${\rm S}_8^0$ reacts with lithium by a two-electron reduction process to form a series of soluble polysulfide intermediates with different chain lengths. Then further polysulfide reduction takes place to produce the solid ${\rm Li}_2{\rm S}$ after full discharge. During charging, the reverse reaction occurs to convert ${\rm Li}_2{\rm S}$ to elemental sulfur ${\rm S}_8^0$ \cite{yin2013lithium}. Fig. \ref{fig:LiS-Batt}(b) depicts a typical discharge voltage profile for Li-S batteries, where two obvious discharge plateaus can be observed. The high plateau, at approximately 2.4 V, corresponds to the formation of ${\rm Li}_2{\rm S}_4$. The reduction of soluble ${\rm Li}_2{\rm S}_4$ into solid ${\rm Li}_2{\rm S}$ occurs in the low plateau at 2.1 V. 

\subsection{Zero-Dimensional Model}

The model equations for the Li-S battery presented here closely follow the derivations in \cite{marinescu2016zero}. This zero-dimensional model considers a two-step electrochemical reaction chain:
\begin{align}
    \label{reac_high}
    & {\rm S}_8^0 + 4{\rm e}^- \longleftrightarrow 2{\rm S}_4^{2-}, \\
    \label{reac_low}
    & {\rm S}_{4}^{2-} + 4{\rm e}^- \longleftrightarrow 2{\rm S}^{2-}\downarrow + {\rm S}_2^{2-},
\end{align}
where each of the two discharge regions is dominated by one of the electrochemical reactions above.

\begin{table}[t]
    \caption{Zero-Dimensional Li-S model Symbol Description} \label{tab:symbols}
	\centering
	\begin{tabular}{  c c c }
	\hline \hline
	Symbols & Description & Units \\ \hline
	$M_{\rm S8}$ & molar mass ${\rm S}_8^0$  & [g/mol]\\
    $n_{\rm S8}$,$n_{\rm S4}$,$n_{\rm S2}$,$n_{\rm S}$ & number of S atoms in polysulfide & [-]\\
    ${n_{\rm e}}$ & electron number per reaction & [-]\\
    $F$ & Faraday's constant & [C/mol]\\
    $R$ & gas constant & [J/K/mol]\\
    $T$ & temperature & [K]\\
    $\rho_S$ & density of precipitated sulfur & [g/L]\\
    $k_s$ & shuttle constant & [s$^{-1}$]\\
    $k_p$ & precipitation rate & [s$^{-1}$]\\
    $S_*^{2-}$ & $S^{2-}$ saturation mass & [g]\\
    $E_H^0$ & standard potential for high plateau (H) & [V]\\
    $E_L^0$ & standard potential for low plateau (L) & [V]\\
    $i_{H,0}$ & exchange current density H & [A/m$^{2}$]\\
    $i_{L,0}$ & exchange current density L & [A/m$^{2}$]\\
    $I$ & Applied current & [A]\\
    $f_H$ & dimensionality factor H & [g L/mol]\\
    $f_L$ & dimensionality factor L & [g$^{2}\cdot$ L$^{2}$/mol]\\
    $a_r$ & active reaction area & [m$^{2}$]\\
    $v$ & electrolyte volume per cell & [L]\\
    $\eta_H$,$\eta_L$ & surface overpotentials &[V]\\ \hline \hline
	\end{tabular}
\end{table}

As a zero-dimensional model, only reactions that occur at the cathode side have been considered and the impact of mass transfer has been neglected. The ``shuttle effects'' of high order polysulfides and the precipitation of lithium sulfide are modelled via shuttle constant ($k_s$) and precipitation rate ($k_p$), respectively. The following dynamical equation describes the time evolution for the various sulfur species in the system,
\begin{equation}
    \label{x_dot}
    \dot{x}(t) = f(x(t),z(t)),
\end{equation}
where $x = [ x_1 \quad x_2 \quad x_3 \quad x_4]^\top \in \mathbb{R}^{n_x}$, $n_x = 4$ is the differential state vector representing the amount of sulfur species for ${\rm S}_8^0$, ${\rm S}_4^{2-}$, ${\rm S}^{2-}$, and ${\rm S_p}$, respectively, and $z = [i_H \quad i_L]^\top \in \mathbb{R}^{n_z}$, $n_z = 2$ is the algebraic state vector representing reaction currents related to the two respective electrochemical reactions \eqref{reac_high} and \eqref{reac_low}. In the present model, it is assumed that the entire sulfur mass is in the form of dissolved ${\rm S}_8^0$ in a fully charged cell. The nonlinear function $f$ is given by
\begin{align}
    \label{f}
    f(x(t),z(t)) = \begin{bmatrix}
    \displaystyle -\frac{n_{\rm S8}M_{\rm S8}}{n_{\rm e}F}i_H - k_s x_1 \\
    \displaystyle \frac{n_{\rm S8}M_{\rm S8}}{n_{\rm e}F}i_H + k_s x_1 - \frac{n_{\rm S4}M_{\rm S8}}{n_{\rm e}F}i_L \\ 
    \displaystyle \frac{2 n_{\rm S}M_{\rm S8}}{n_{\rm e}F}i_L - \frac{1}{v \rho_S} k_p x_4(x_3-S_*^{2-}) \\
    \displaystyle \frac{1}{v \rho_S} k_p x_4(x_3-S_*^{2-})
    \end{bmatrix}.
\end{align}

The equilibrium potentials for the reactions \eqref{reac_high}-\eqref{reac_low} are described by the Nernst equations
\begin{align}
    \label{EH}
    E_H = {} & E_H^0 + \frac{RT}{4F}\ln\Big(f_H \frac{x_1}{x^2_2}\Big), \\
    \label{EL}
    E_L = {} & E_L^0 + \frac{RT}{4F}\ln\Big(f_L \frac{x_2^2}{x_3^2(x_3+x_4)}\Big).
\end{align}

\begin{remark}
Under the assumption of mass conservation between ${\rm S_2^{2-}}$ and the sum of ${\rm S^{2-}}$ and ${\rm S_p}$, i.e., $m({\rm S^{2-}}) = m({\rm S^{2-}}) + m({\rm S_p})$, the time evolution of ${\rm S^{2-}}$ has been ignored. Specifically, the five-order state equations presented in Eqs. (8a)-(8e) in \cite{marinescu2016zero} is equivalently reduced to the four-order system \eqref{x_dot}-\eqref{f} in this work. Furthermore, this model reduction produces an observable DAE system in the linear sense as detailed in Section \ref{s:observability} below.
\end{remark}

When current is present, the battery is in a non-equilibrium state. Under this condition, the currents associated with the two electrochemical reactions \eqref{reac_high} and \eqref{reac_low} are given by the Butler-Volmer equations:
\begin{align}
    \label{iH}
    i_H = {} & 2 i_{H,0} a_r \sinh\Big(\frac{n_{\rm e} F \eta_H}{2RT}\Big), \\
    \label{iL}
    i_L = {} & 2 i_{L,0} a_r \sinh\Big(\frac{n_{\rm e} F \eta_L}{2RT}\Big).
\end{align}
A non-zero surface overpotential, denoted by $\eta_H, \eta_L$, is the driving force for a reaction to occur and it is given by the difference between the voltage of the cell and the reaction Nernst potential \cite{marinescu2016zero},
\begin{align}
    \label{etaH}
    \eta_H = V - E_H,\\
    \label{etaL}
    \eta_L = V - E_L.
\end{align}
The output of the system can then be written as
\begin{align}
    \label{voltage}
    y(t) = {} & h(x(t),z(t)),
\end{align}
where 
\begin{align}
    \label{out_stack}
    &h(x(t),z(t)) = \nonumber\\
    &\begin{bmatrix} \displaystyle E_H^0 + \frac{RT}{4F}\ln\big(f_H \frac{x_1}{x^2_2}\big) + \frac{2RT}{n_{\rm e}F}\sinh^{-1}\big(\frac{i_H}{2i_{H,0}a_r}\big) \\[0.3cm]
    \displaystyle E_L^0 + \frac{RT}{4F}\ln\big(f_L \frac{x_2^2}{x_3^2(x_3+x_4)}\big) + \frac{2RT}{n_{\rm e}F}\sinh^{-1}\big(\frac{i_L}{2i_{L,0}a_r}\big)\end{bmatrix}.
\end{align}
The output vector $y(t) = [y_1(t) \quad y_2(t)]^\top$ represents the voltage measurement of the Li-S cell computed from the high and low voltage plateau kinetics, respectively. Namely, $y_1(t) = V(t) = \eta_H(t) + E_H(t)$ and $y_2(t) = V(t) = \eta_L(t) + E_L(t)$, which are (the same) measured signals.

Finally, the measured cell current $I$ is the summation of currents from the two reactions, i.e. 
\begin{equation}
    \label{I}
    I = i_H + i_L.
\end{equation}

\subsection{Differential-Algebraic System}

The dynamical equations \eqref{x_dot}-\eqref{f} as well as the algebraic constraints \eqref{EH}-\eqref{I} can be arranged in the following compact state-space form as a nonlinear differential-algebraic system,
\begin{align}
    \label{diff}
    \dot{x}(t) = {} & f(x(t),z(t)), \\
    \label{alg}
    0 = {} & g(x(t),z(t),u(t)), \\
    \label{DAE_out}
    y(t) = {} & h(x(t),z(t)),
\end{align}
with function $g(x(t),z(t),u(t))$ given by
\begin{align}
    \label{g}
    g(x,z,u) = \begin{bmatrix} g_1(x,z,u) & g_2(x,z,u) \end{bmatrix}^\top,
\end{align}
with
\begin{align}
    g_1 = {} & i_H + i_L - I, \\
    g_2 = {} & \displaystyle E_H^0 + \frac{RT}{4F}\ln\Big(f_H \frac{x_1}{x^2_2}\Big) - E_L^0 - \frac{RT}{4F}\ln\Big(f_L \frac{x_2^2}{x_3^2(x_3+x_4)}\Big) \nonumber \\
    \displaystyle & + \frac{2RT}{n_{\rm e}F}\sinh^{-1}\Big(\frac{i_H}{2i_{H,0}a_r}\Big) - \frac{2RT}{n_{\rm e}F}\sinh^{-1}\Big(\frac{i_L}{2i_{L,0}a_r}\Big).
\end{align}
We have substituted \eqref{EH}-\eqref{iL} into \eqref{etaH}-\eqref{etaL} to form \eqref{g}. System \eqref{diff}-\eqref{DAE_out} can be conveniently verified to be a semi-explicit DAE of index 1 as $\partial g/ \partial z$ has full rank (invertible) \cite{brenan1995numerical}. It is also worth highlighting that in some battery applications, e.g., \cite{zhang2020parallel}, function $g$ is linear in $z$ such that under suitable conditions we can explicitly solve for $z$ in terms of $x$ and substitute it back into \eqref{diff} to form a reduced ODE system. This is not applicable in the Li-S battery system because function $g$ is highly nonlinear with respect to both $z$ and $x$, prohibiting a closed form solution of constraint \eqref{alg}.

\section{Observability Analysis} \label{s:observability}

Prior to state observer design, it is crucial to conduct an observability analysis to confirm that the entire state-space, including both the differential and algebraic states, can be reconstructed from input-output data. In this section, we mathematically analyze the local observability of the nonlinear differential-algebraic system \eqref{diff}-\eqref{DAE_out}.

Let $w = [x \quad z]^\top$ be the augmented state. In order to study the observability of the nonlinear DAE system \eqref{diff}-\eqref{DAE_out}, we linearize the system around an equilibrium point $w = w_0$ and verify the observability conditions for the linearized system. If the linearized system is observable at $w = w_0$, then the nonlinear system is locally observable. However, it is further noted that the observability results from linearizing the nonlinear system is only sufficient, i.e., no conclusion can be drawn for the nonlinear system if the linearized system is not observable \cite{zhang2020parallel}. The linearized model of \eqref{diff}-\eqref{DAE_out} is given by
\begin{align}
    \label{x_lin}
    E \dot{w}(t) = {} & A w(t) + B u(t), \\
    \label{y_lin}
    y = {} & C w(t),
\end{align}
where the state matrix $A \in \mathbb{R}^{(n_x+n_z)\times(n_x+n_z)}$ and output matrix $C \in \mathbb{R}^{2\times(n_x+n_z)}$ are given by
\begin{align}
E = & \begin{bmatrix} I_{n_x \times n_x} & \mathbf{0}_{n_x \times n_z} \\ \mathbf{0}_{n_z \times n_x} & \mathbf{0}_{n_z \times n_z} \end{bmatrix}, \quad
A = \begin{bmatrix} \displaystyle \frac{\partial f}{\partial x} &\displaystyle \frac{\partial f}{\partial z} \\\displaystyle \frac{\partial g}{\partial x} &\displaystyle \frac{\partial g}{\partial z} \end{bmatrix}_{w = w_0}, \nonumber \\
C = & \begin{bmatrix}\displaystyle \frac{\partial h}{\partial x} &\displaystyle \frac{\partial h}{\partial z} \end{bmatrix}_{w = w_0}.
\end{align}

\begin{figure}[t]
	\centering
	\includegraphics[trim = 4.5mm 4mm 9mm 2mm, clip, width=\linewidth]{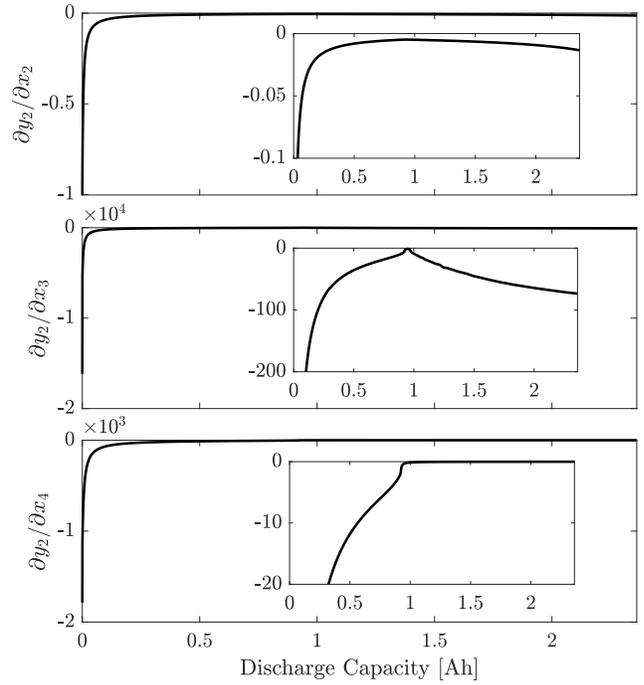}
	\caption{The sensitivities of low plateau voltage with respect to system states, simulated using a constant discharge current at 1.7 A.}
	\label{fig:sensitivity}
\end{figure}

Let us now introduce the definition of complete observability (C-observability) for a linear DAE system \cite{duan2010analysis}.

\begin{theorem} [\cite{duan2010analysis}]
\label{thmcobs}
The regular linear differential-algebraic system \eqref{x_lin}-\eqref{y_lin} is C-observable if and only if the following two conditions hold:
\begin{enumerate}
    \item[C.1] $\mathrm{rank}\left\{ [ E^\top, C^\top ]^\top \right\} = n_x + n_z$;
    \item[C.2] $\mathrm{rank}\left\{ [ (s E - A)^\top, C^\top ]^\top \right\} = n_x + n_z, \ \forall s \in \mathbb{C}$.
\end{enumerate}
\end{theorem}

Condition C.1 verifies C-observability of the algebraic subsystem (fast subsystem) while C.2 involves the dynamic subsystem (slow subsystem). Condition C.1 is satisfied if $\partial h/\partial z$ has column rank $n_z$, since the identity sub-matrix in $E$ already provides a column rank of $n_x$. This can be easily verified from \eqref{out_stack}.

The verification of condition C.2 requires the numerical computation of the generalized eigenvalues of the pair $(E,A)$. Although condition C.2 has to be validated against all $s$ in complex domain, it is automatically verified when $s$ is not one of the generalized eigenvalues of pair $(E,A)$. The sensitivity of the equilibrium potential with respect to system states are demonstrated in Fig. \ref{fig:sensitivity}, which is simulated using a constant discharge rate at 1.7 A. During discharge at low plateau, in which the transition from high to low occurs at around 1 Ah position, the sensitivity values are significantly smaller than those from high plateau, indicating relatively weak observability. 

\section{Observer Design} \label{s:obs}

In this section, an EKF approach for nonlinear DAE systems, similar to the UKF reported in 
\cite{Couto-2018}, is used for estimation in our system. This algorithm applies to measured outputs that are functions of both differential and algebraic state variables. The standard EKF algorithm for ODE systems, however, can only be applied when the differential 
states are decoupled from the algebraic ones. Then the algebraic states can be computed as implicit solutions to the nonlinear algebraic constraints at each time step.

The nonlinear DAE system \eqref{diff}-\eqref{DAE_out} is transformed into discrete-time domain to facilitate the implementation of the EKF for DAEs,
\begin{align}
    \label{xk_dis}
    x_{k+1} = {} & x_{k} + \Delta t \cdot f(x_k, z_k) + \mu_{k} \\
    \label{alg_dis}
    0 = {} & g(x_k, z_k, u_k), \\ 
    \label{y_dis}
    y_{k+1} = {} & h\left(x_{k+1}, z_{k+1}\right)+\nu_{k+1}
\end{align}
where $\Delta t$ is the sampling time, $x_{k+1}$ and $z_{k+1}$ are the discretized differential and algebraic states at time $t=(k+1) \Delta t$, respectively, and $\mu_{k+1}$ and $\nu_{k+1}$ are assumed to be stationary, zero-mean and Gaussian white noise processes with covariance matrices $Q$ and $R$, respectively. At time step $k$, the differential state is propagated in time to time step $(k+1)$ according to \eqref{xk_dis} using the current differential and algebraic values, $x_k$ and $z_k$. Once the differential states at time step $(k+1)$ is obtained, the algebraic equation \eqref{alg_dis} is solved numerically for $z_{k+1}$. This process is repeated to simultaneously propagate differential and algebraic states in time.

\begin{algorithm}[t]
\SetAlgoLined
\textbf{Inputs:} $u_k$, $y_k$, $k = 1,2, \cdots$ \\
\textbf{Outputs:} $\hat{x}_k$, $\hat{z}_k$, $k = 1,2, \cdots$ \\
\begin{itemize}[leftmargin=*]
    \item At time step $k$, the (consistent) estimates of the algebraic states are calculated using the algebraic equations of DAE system to satisfy the algebraic constraints:
    \begin{align}
        g(\hat{x}_{k}, \hat{z}_{k}, u_k) = 0. \nonumber
    \end{align}
    \item Given the up-to-date estimates $\hat x_{k}$ and consistent algebraic state estimates $\hat{z}_{k}$, the differential state estimates are propagated forward in time using the nonlinear discrete-time model and corrected through output error injection as
    \begin{equation}
        \hat{x}_{k+1} = (\hat{x}_{k} + \Delta t \cdot f(\hat{x}_{k},\hat{z}_{k})) 
        + K_{k}\left(y_{k}-h(\hat{x}_{k},\hat{z}_{k})\right),
        \nonumber
    \end{equation}
    in which the forward Euler method has been employed.
    \item The computation of the covariance matrix of the differential state estimation error is given by
    \begin{align}
        P_{k+1} = 
        F_k P_{k} F_k^\top + Q - K_k \left( H_K P_k H_k^\top + R \right) K_k^\top, \nonumber
    \end{align}
    and the Kalman gain matrix is
    \begin{align}
        K_{k} = F_k P_{k} H_{k}^\top\left(H_{k} P_{k} H_{k}^\top + R\right)^{-1}, \nonumber
    \end{align}
    where $F_{k}$ and $H_{k}$ are the linearized state and output equations with respect to the differential state evaluated at $\hat{x}_{k}$,
    \begin{align}
        F_{k} = \frac{\partial f}{\partial x}\bigg|_{\hat{x}_{k}}, \quad
        H_{k} = \frac{\partial h}{\partial x}\bigg|_{\hat{x}_{k}}.  \nonumber 
    \end{align}
\end{itemize}
\caption{EKF for Nonlinear DAEs}
\label{EKF_Algorithm}
\end{algorithm}

The EKF mathematics are reported in Algorithm \ref{EKF_Algorithm}. Essentially, the algorithm first computes algebraic states that are consistent with the DAE, i.e. they satisfy the nonlinear algebraic equations. 
Then, both the differential and algebraic states at time $k$ are used to predict the differential state at the next time instant through the nonlinear state function as well as correct it via output error injection. 
The linearized ODE model is subsequently used for the covariance propagation of the differential states, followed by the computation of the gain matrix using the covariance matrix and the linearized model matrices. 
It should be noted that this algorithm only performs estimation update for the differential states using the classical Kalman filter approach \cite{Goodwin-1984}, whilst the algebraic state estimates are updated at each time step by solving the nonlinear algebraic constraints.


\begin{figure}[t]
	\centering
	\includegraphics[trim = 3mm 5mm 12mm 8mm, clip, width=\linewidth]{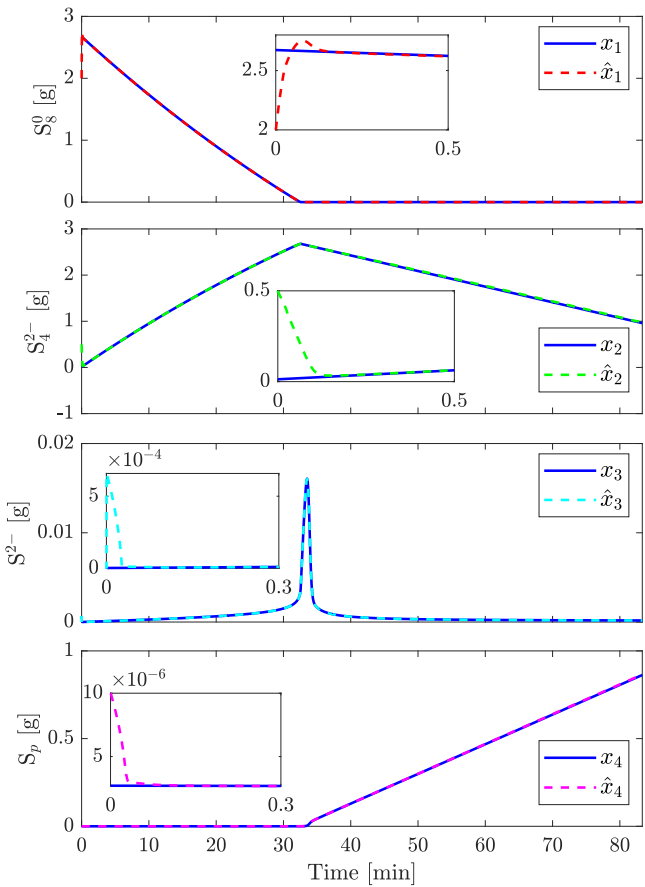}
	\caption{The estimation performance for the amount of sulfur species in discharge process.}
	\label{fig:diff}
\end{figure}

\section{Simulation Results} \label{s:sim}

\begin{figure}[t]
	\centering
	\includegraphics[trim = 6mm 4.5mm 12mm 7mm, clip, width=\linewidth]{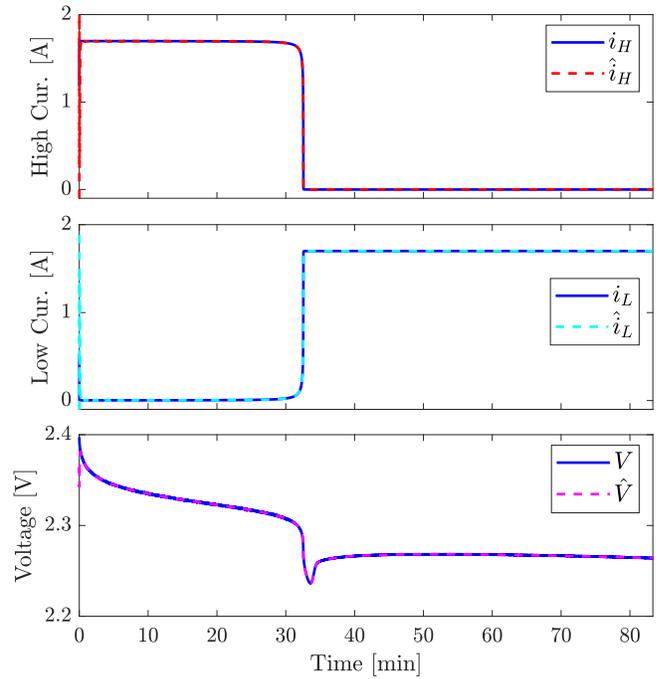}
	\caption{The estimation performance for the algebraic states ($i_H$ and $i_L$) as well as voltage in discharge process.}
	\label{fig:Alg}
\end{figure}

In this section, we present studies on simulation to demonstrate the performance of the proposed EKF-based estimation scheme for nonlinear DAE systems \eqref{diff}-\eqref{DAE_out}. The parameter values for plant model and estimator are obtained from \cite{marinescu2016zero}. 
We apply a constant 1.7 A discharge current to the plant model from a fully-charged state for 5000 seconds (approx 83 minutes). At fully charged state, the entire sulfur mass is in the form of dissolved S$_8^0$. At high plateau, the dissolved S$_8^0$ is consumed to form S$_4^{2-}$. The low plateau is reached when S$_8^0$ has been entirely consumed, as demonstrated in Fig. \ref{fig:diff}. The estimation is conducted by using only the applied current and terminal voltage measurements, where the voltage signal was corrupted with a zero-mean Gaussian noise sequence of $10^{-4}$ mV variance. 
The actual initial conditions of the states in the plant model are $w_0 = [2.6730 \quad 0.0128 \quad 8.9339\times 10^{-7} \quad 2.7\times 10^{-6} \quad 1.7 \quad 0]^{\top}$, whereas the EKF is initialized with  $\hat w_0 =[2 \quad 0.5 \quad 10^{-5} \quad 10^{-5} \quad -0.1838 \quad 1.8838]^{\top}$. Fig. \ref{fig:diff} presents the estimates for the mass of various sulfur species as well as their simulated values from the plant model. The estimation of algebraic states ($i_H$ and $i_L$) as well as the voltage ($V$) are plotted against their true values in Fig. \ref{fig:Alg}. By tuning the EKF with the following parameters,
\begin{align}
P_{0} & = \text{diag}([2.5\times 10^{-6} \quad 10^{-7} \quad 10^{-11} \quad 4.895\times 10^{-15}]),\nonumber\\ 
Q_{k+1} & = \text{diag}([10^{-12} \quad 10^{-12} \quad 10^{-15} \quad 10^{-21}]),\nonumber \\ 
R_{k+1} & = \text{diag}([10^{-7} \quad 10^{-7}]),\nonumber
\end{align}
the estimates effectively converge to their true values (within approximately 0.5 min) from large initial estimation errors, according to the design procedures presented in Section \ref{s:obs}.  It is worth noting that during low plateau, the state estimates slightly diverge from the plant model simulated signals, which confirms our conclusion in Section \ref{s:observability} that the states are weakly observable from the system input-output data. Although not reported in detail here, we have additionally tested how much initialization error this EKF can withstand until the estimates diverge. Our experiments show that the initial estimates must be physically unreasonable before the algorithm diverges, thereby demonstrating significant robustness to state initialization error.

\section{Conclusion} \label{s:conclusion}
This paper explores the potential use of the reduced-order electrochemical model for state estimation of Li-S batteries, which have 10X the theoretical specific energy of Li-ion batteries. A state observer based on a zero-dimensional electrochemical model has been presented. Observability of the linearized DAE system has been studied. The analysis shows the states are indeed locally observable, but observability is relatively weak in the low plateau region. An extend Kalman filter-based algorithm is then adopted to estimate the amount of sulfur species, as well as the reaction kinetics inside the cell. The accuracy of the proposed estimation approach is demonstrated in simulation. Real-time monitoring of the electrochemical state information enables i) a further understanding of the electrochemical mechanisms inside the cells, and ii) high-performance control and operation in advanced BMSs for practical applications, including electric aircraft and long-haul trucks. This estimation scheme will be further validated using experimental data of Li-S batteries. Additionally, approaches to estimate the SOC and SOH of Li-S cells based on this reduced-order electrochemical model will be studied in the future.

\bibliographystyle{ieeetr}
\bibliography{Reference} 

\end{document}